\documentclass[twocolumn,showpacs,preprintnumbers,amsmath,amssymb]{revtex4}

\def\XXint#1#2#3{{\setbox0=\hbox{$#1{#2#3}{\int}$}
     \vcenter{\hbox{$#2#3$}}\kern-.5\wd0}}

\def\fakebold#1{\relax\ifvmode\leavevmode\fi%
\ifmmode%
\setbox0=\hbox{$#1$}%
\else%
\setbox0=\hbox{#1}%
\fi%
\kern-.02em\copy0 \kern-\wd0%
\kern .04em\copy0 \kern-\wd0%
\kern-.0125em\raise.02em\box0%
}%

\usepackage{graphicx}
\usepackage{dcolumn}
\usepackage{bm}

\begin{document}



\title{Temperature dependent effective mass renormalization in a
  Coulomb Fermi liquid}

\author{Ying Zhang} \author{S. Das Sarma} \affiliation{Condensed
  Matter Theory Center, Department of Physics, University of Maryland,
  College Park, MD 20742-4111}

\date{\today}

\begin{abstract}
We calculate numerically the quasiparticle effective mass ($m^*$)
renormalization as a function of temperature and electron density in
two- and three-dimensional electron systems with long-range Coulomb
interaction. In two dimensions, the leading temperature correction is
linear and positive with the slope being a universal density
independent number in the high density limit. We predict an
enhancement of the effective mass at low temperatures and a
non-monotonic temperature dependence at higher temperatures ($T/T_F
\sim 0.1$) with the peak shifting toward higher temperatures as
density decreases. In three dimensions, we find that the effective
mass temperature dependence is nonlinear and non-universal, and
depends on the electron density in a complicated way. At very high
densities, the leading correction is positive, while at lower
densities it changes sign and the effective mass decreases
monotonically from its zero temperature value with increasing
temperature.
\end{abstract}


\maketitle


\section{Introduction}
In the Fermi liquid theory the interacting electron system is composed
of weakly interacting quasiparticles at low energies with long
quasiparticle lifetimes. The effective mass of a quasiparticle, which
can be viewed as the bare mass of a free electron being renormalized
by electron-electron interactions, is an important and fundamental
Fermi liquid parameter.  For decades theorists have been exploring the
effective mass renormalization in two- and three- dimensional
interacting electron systems (2DES and 3DES). In spite of this great
deal of theoretical activity concentrating almost entirely on the
density dependence of the effective mass renormalization, the
temperature dependence of the effective mass has not been studied
until very recently. Besides the considerable difficulties involved in
the finite temperature numerical and analytical many-body calculations
in 2DES and 3DES, the reason that this issue has not been addressed
before can also be explained by the fact that the Fermi energy in 3D
metals is typically $10^4$K, and therefore any finite temperature
effects are negligible. In the past decade, however, low density 2DES
have been attracting attention, and several experiments have been
performed to measure the 2D effective mass~\cite{oldexp,exp}.  The
temperature dependence of the quasiparticle effective mass in 2DES is
of considerable experimental interest since the Fermi energy in
realistic 2DES may be $1K$ or lower, which makes the issue of the
temperature dependence of 2D Fermi liquid parameters extremely
important. In addition, the temperature dependence of the Fermi liquid
parameters such as the effective mass is obviously of considerable
fundamental theoretical significance.

The $T=0$ quasiparticle effective mass renormalization in an electron
system interacting through the long-range Coulomb interaction is one
of the oldest many-body problems in theoretical condensed matter
physics, and a number of theoretical calculations of 3D and 2D
electron effective mass have been carried
out~\cite{rice,gellmann,ting,vinter,jalabert,marmorkos,book} in the
literature. In fact, the Coulomb interaction induced electron
effective mass renormalization at $T=0$ is standard text-book
material~\cite{book} in electronic many-body theory.  Essentially all
of these calculations, both analytical and numerical (and both 2D and
3D), are based on the leading-order dynamically screened interaction
one-loop self-energy evaluation (the so-called RPA or `GW' self-energy
approximation) because this approximation is really the only
meaningful nontrivial calculation that can actually be carried out,
and (perhaps more importantly) because this RPA self-energy is
asymptotically exact in the weakly interacting high density regime.
There have been a few finite temperature RPA self-energy calculations
over the years~\cite{chaplik,gq,hu,dassarma} mostly in the context of
low dimensional systems, but none for the temperature dependence of
the effective mass renormalization in interacting electron systems.
Very recently, Chubukov and Maslov~\cite{chubukov} considered the
problem of temperature corrections to the 2D Fermi liquid theory for
the case of a short-ranged interaction. In particular, they showed
that the leading many-body temperature correction is linear in 2D
similar to the results which we reported recently for the long-range
Coulomb interaction in 2DES~\cite{short}.

In the current paper we present a calculation of the density and
temperature dependent effective mass renormalization by the Coulomb
interaction in 2DES and 3DES at arbitrary densities and temperatures
(i.e. {\em not} necessarily restricted to high densities and low
temperatures). We work within the random-phase approximation (RPA), or
equivalently in the ring-diagram approximation for the self-energy,
which gives exact results in the high-density limit ($r_s \ll 1$) but
is known to be qualitatively reliable at relatively low densities as
well. RPA is perhaps the only manageable way to perform any
non-trivial quantitative calculations in electronic many-body systems,
and the finite temperature RPA effective mass renormalization is
certainly a problem of intrinsic interest. In two dimensions, our
numerical results predict a non-monotonic effective mass temperature
dependence. The leading temperature dependence is linear and positive,
with the low-temperature slope being independent of the electron
density in the high density limit. The temperature at which the
effective mass is maximum at a particular density moves toward higher
temperatures as density decreases. In three dimensions, we find that
the effective mass temperature dependence is non-universal and depends
on the electron density in a complicated way. At very high densities,
the leading correction is positive, while at lower densities it
changes sign and decreases monotonically from its zero temperature
value. This is in contrast to the 2D results where the effective mass
always increases (linearly) with temperature at low temperatures, and
then decreases with temperature beyond a density-dependent
characteristic temperature. We find the 3D temperature correction to
the effective mass to be nonlinear in contrast to our 2D results.

We express the quasiparticle effective mass $m^*(n, T) \equiv m^*(r_s,
T/T_F)$ in units of the bare band mass $m$ (which is, by definition, a
constant) and present our results as a function of the usual
dimensional interaction parameter $r_s$ (the average inter-electron
separation measured in the units of Bohr radius) and the dimensionless
temperature $T/T_F$ where $T_F = E_F / k_B$ is the Fermi temperature.
Note that $r_s \propto n^{-1/2}~(n^{-1/3})$ in 2D (3D) systems, and
$T_F \propto n~(n^{2/3})$ in 2D (3D) systems, where $n$ is the
appropriate 2D (per unit area) or 3D (per unit volume) electron
density. Note that the dimensionless interaction and temperature
parameters $r_s$ and $T/T_F$ are {\em not} independent parameters
since they both depend on the electron density. We also note that $r_s
\ll 1$ (high-density) and $r_s \gg 1$ (low density) limits are
respectively the weak- and the strong-interaction limits of the
electron system (at $T=0$), and $T/T_F \ll 1$ and $T/T_F \gg 1$ are
respectively the low-temperature (quantum) and the high-temperature
(classical) limits. We consider the electron system to be a uniform
jellium system with the noninteracting kinetic energy dispersion being
the usual parabolic dispersion. We use $\hbar = k_B = 1$ throughout.

The structure of our paper is as follows: In section~\ref{sec:form} we
provide the formalism which we will use in this paper. In
section~\ref{sec:method} we explain in detail the numerical method we
are using in the effective mass calculations. In
section~\ref{sec:results} we present all our numerical results for 2D
and 3D effective mass, comparing to analytical results in the high
density limit. In section~\ref{sec:PPA} we discuss a special
approximation method, the plasmon-pole approximation, and present our
effective mass results using this method. In section~\ref{sec:decay},
we calculate the imaginary self-energy of quasiparticles and discuss
the validity of the quasiparticle approximation at finite
temperatures. We provide a conclusion and discussion of our results in
section~\ref{sec:con}.


\section{Formalism}
\label{sec:form}

In this section we give the theoretical formalism, the basic
equations, and the notations which will be used throughout the paper.

\subsection{Effective mass}

In a system of interacting fermions the retarded Green's function can
be written as
\begin{equation}
\label{Gr}
G_R ({\bf k}, \omega) 
= {1 \over\omega  -\epsilon_0({\bf k})+\mu + \Sigma({\bf k},\omega)}, 
\end{equation}
where $\epsilon_0({\bf k}) = k^2 /2m$ is the spectrum of
non-interacting fermions, $\mu$ is the chemical potential, and
$\Sigma({\bf k},\omega)$ is the quasiparticle self-energy, the
imaginary (real) part of which determines the lifetime (effective
mass) of the quasiparticle. The quasiparticle energy can be obtained
by solving the Dyson's equation~\cite{book}
\begin{equation}
\label{eq:dyson}
\epsilon({\bf k}) =  \epsilon_0({\bf k})+ {\rm
  Re\,}\Sigma({\bf k},\epsilon({\bf k})).
\end{equation}
The quasiparticle effective mass can be written by definition as
\begin{eqnarray}
\label{eq:offshell}
{m^* \over m} &=& {m \over k} {d \over d k} \epsilon({\bf k}) 
\Big|_{k=k_F} \nonumber \\
&=& {1 - {\partial \over \partial \omega} {\rm
    Re\,} \Sigma\left({\bf k}, \omega\right)
    \over 1 + {m \over k} {\partial \over \partial p} {\rm Re\,} 
    \Sigma({\bf k}, \omega)} \Big|_{k = k_F, \omega = 0}. 
\end{eqnarray}
Note that in the above equation $\omega = 0$ is measured from the
renormalized chemical potential $\mu^*$, which is given by
\begin{equation}
\label{eq:mu}
\mu^* = \mu + {\rm Re} \Sigma(k_F, 0).
\end{equation}

All the above equations are exact, while the RPA approximation for
$\Sigma({\bf k}, \omega)$ that we are going to use is the first order
perturbation theory in the dynamically screened interaction. There has
been extensive discussion~\cite{rice,ting,book,diverge} on whether one
should use exact Eq.(\ref{eq:offshell}) for calculating the effective
mass or it is more consistent to use the so-called on-shell
approximation, keeping only the first order interaction terms in the
expression for the effective mass (since $\Sigma$ is calculated only
to first order in the dynamically screened interaction):
\begin{equation}
\label{eq:onshellm}
{m^* \over m} = {1 \over 
1 +  {m \over k} {d \over d k} 
\Sigma \left({\bf k}, \xi_{\bf k} \right)\big|_{k=k_F} },
\end{equation} 
where $\xi_{\bf k} = k^2 /(2m) - \mu$. Note that all the quantities on
the right side of Eq.~(\ref{eq:onshellm}) are in the leading order in
effective interaction. There are compelling arguments in favor of the
latter choice: the on-shell approximation is believed to be more
accurate as it effectively accounts for some higher order diagrams and
satisfies the Ward identity. We have extensively discussed this issue
elsewhere~\cite{diverge}.

Obviously, the two equations for calculating $m^*$ are identical in
the high-density limit $r_s \ll 1$. However, in the region of $r_s >
1$, they give very different results. In what follows, we use
Eq.~(\ref{eq:onshellm}) for all the numerical results shown in this
paper because we believe the on-shell approximation to be the superior
one in our case. Both formulae give similar temperature dependence for
$m^*(T)$. The main qualitative results of the paper are insensitive to
the choice of the on-shell or off-shell formula for the effective
mass.

\subsection{Self-energy in the RPA approximation.}

\begin{figure}[htbp]
\centering \includegraphics[width=2in]{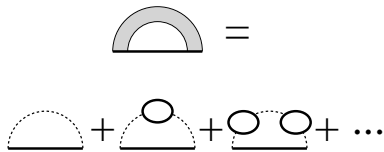}
  \caption{Feynman diagram for RPA self-energy calculation. Solid lines
    denote the free electron Green's function and the dashed lines the
    bare Coulomb potential.}
\label{fig1}
\end{figure}

Within RPA, the finite temperature electron self-energy can be
expressed in terms of the Feynman diagrams shown in Fig.~\ref{fig1},
and can be written in the Matsubara formalism as~\cite{book}:
\begin{equation}
\label{eq:EMats}
\Sigma ({\bf k}, \nu_n) = - T \sum\limits_{\omega_m} {\cal
  G}({\bf k}-{\bf q}, \nu_n - \omega_m) {\cal D}({\bf q},\omega_m), 
\end{equation}
where $ \nu_n = \pi (2 n +1) T$ is the fermion Matsubara frequency,
$\omega_m = 2 \pi m T$ is the boson Matsubara frequency with $n$ and
$m$ integers, and $T$ the temperature. The function ${\cal D}({\bf
  q},\omega_m)$ denotes the coupling to a collective mode (phonon,
plasmon, electron-hole excitation, etc.), {\em i.e.}, ${\cal D}$ is
the bosonic propagator for the effective interaction. In our case, the
function is the dynamically screened Coulomb interaction given by the
sum of the ring or bubble diagrams:
\begin{equation}
\label{eq:scr_Coulomb}
{\cal D}({\bf q},\omega_m) = {v_0({\bf q})  \over 1 + v_0({\bf q})
  \mbox{\Large $\pi$} ({\bf q},\omega_m) },
\end{equation} 
where $v_0({\bf q})$ is the bare Coulomb interaction and $\mbox{\Large
  $\pi$}({\bf q},\omega_m)$ is the (bare) polarization operator, which
is defined as
\begin{eqnarray}
\label{eq:PMats}
\mbox{\Large $\pi$} \left({\bf q},\omega_m\right) &=& 
2 \sum\limits_{\nu_n} \int {d ^d {\bf p} \over \left(2
  \pi \right)^d} {\cal G}^{(0)} ({\bf p}, \nu_n) 
\nonumber \\
&&~~~~~~~~~~~\cdot {\cal G}^{(0)}({\bf p + q}, \nu_n + \omega_m), 
\end{eqnarray}
where $d$ is the dimension of the system and ``$(0)$'' denotes the
non-interacting system. We mention that Eqs.~(\ref{eq:scr_Coulomb})
and (\ref{eq:PMats}) together form what is called the RPA for an
electron gas, where the bare Coulomb interaction is dynamically
screened by the electron dielectric function, which is formed from the
infinite series of the polarization bubbles. The corresponding
electron self-energy, obtained in the leading-order expansion in the
dynamically screened interaction ${\cal D}$, is conventionally called
the RPA self-energy approximation, although the ``dynamical
Hatree-Fock'' approximation or the ``Ring Diagram Approximation'' may
be a more appropriate terminology.

For calculations, it is more convenient to use the self-energy defined
as a function of the real frequency $\omega$ rather than the Matsubara
one. Using the standard procedure of analytic continuation, one
obtains the following expression for the analytically continued
self-energy:
\begin{eqnarray}
\label{eq:Sig_gen}
&&\Sigma^R \left( {\bf k},\omega \right) = 
-\int {d^d {\bf q} \over (2 \pi)^d}
\int\limits_{-\infty}^{+\infty} {d \nu \over 2 \pi}
\Big[ \nonumber \\
&&~~~~~~~{\rm Im\,} G_R^{(0)}
\left({\bf k} - {\bf q}, \nu + \omega \right) 
D_R\left({\bf q}, -\nu \right) \tanh ( {\nu +
  \omega \over 2 T} ) \nonumber \\ 
&&~~~+ G_R^{(0)}
 \left({\bf k} - {\bf q}, \nu + \omega \right)
{\rm Im\,} D_R\left({\bf q}, \nu \right) \coth ( {\nu
  \over 2 T} ) \Big], 
\end{eqnarray}
where functions labeled with index ``R'' are retarded functions, {\em
  i.e.} functions analytical in the upper half-planes of the complex
frequency. The corresponding effective interaction can be written as:
\begin{equation}
\label{eq:scr_Coulomb1}
 D_R({\bf q},\omega ) = {v({\bf q}) \over 1 + v({\bf q})
 \Pi_R ({\bf q},\omega  ) },
\end{equation} 
where the retarded polarizability can be obtained from
Eq.(\ref{eq:PMats}) using the following identities:
\begin{equation}
\label{PRA}
\Pi_R ({\bf q},\omega)=\mbox{\Large $\pi$}
({\bf q},i \omega_n \to \omega + i \eta), 
\end{equation}
where $\eta$ is a real infinitesimal positive number.

Note that we will almost always use retarded quantities unless
otherwise stated. Thus without causing any confusion, we can drop the
superscript ``R''.


\subsection{Effective interaction}
\label{sec:form:eff}
The next step toward deriving the renormalization of mass is to obtain
expressions for the effective coupling $D({\bf q},\omega)$. We use the
long-range bare Coulomb interaction to get
\begin{eqnarray}
\label{eq:bareV}
v_0^{(2D)}(q) &=& {2 \pi e^2 \over q}, \nonumber \\
v_0^{(3D)}(q) &=& {4 \pi e^2 \over q^2},
\end{eqnarray}
and the effective interaction
\begin{equation}
\label{eq:scr_Coulomb2}
 D({\bf q},\omega) = {v_0({\bf q})  \over 1 + v_0({\bf q}) \Pi
  ({\bf q},\omega) } = {v_0({\bf q})  \over \epsilon({\bf q},
  \omega)}, 
\end{equation} 
where $\epsilon({\bf q}, \omega) \equiv 1 + v_0 \Pi$ is the RPA
dynamical dielectric function. In the RPA the full polarizability is
approximated by the bare polarizability as in Eq.~(\ref{eq:PMats}),
which is just the bare bubble diagram.

Analytical properties of the propagator $\Pi_0$ (where $\Pi_0$ denotes
the $T=0$ form for the bare polarizability with $\Pi$ denotes the
finite $T$ bare polarizability) are non-trivial even at zero
temperature. The zero temperature polarization for 2DES and 3DES are
well-known and shown below. For 2D $T=0$ case we have
\begin{eqnarray}
\label{eq:2DP}
\Pi^{\mbox{2D}}_0 (q, \omega, \mu) 
= -{m \over \pi} + {m^2 \over \pi q^2} 
\Bigg[ \sqrt{ (\omega + {q^2 \over 2m} )^2 - {2 \mu q^2 \over 2m} } 
\nonumber \\
- \sqrt{ (\omega - {q^2 \over 2m} )^2 
- {2 \mu q^2 \over 2m} } \Bigg],
\end{eqnarray}
where $\mu$ is the chemical potential, the frequency $\omega$ can be
any complex number, and the branch cut of the square roots are taken
so that the imaginary part is positive. For 3D ($T=0$) case we have
\begin{eqnarray}
\label{eq:2DPT}
\Pi^{\mbox{3D}}_0 (q, \omega, \mu) 
&=& {k_\mu m \over 2 \pi^2 q^2} \Bigg\{ 1 \nonumber \\
&& \!\!\!\!\!\!\!\!\!\!\!\!\!\!\!\!\!\!\!\!\!\!\!\!\!\!\!\!\!\!\!\!\!\!\!\!\!\!\!\!\!\!\!\!\! 
+ {m^2 \over 2 k_\mu q^3} 
\left[4 \mu \epsilon_q - (\epsilon_q + \omega)^2 \right]
\ln \left({\epsilon_q + q v_\mu + \omega 
  \over \epsilon_q + q v_\mu + \omega} \right) \nonumber \\
&& \!\!\!\!\!\!\!\!\!\!\!\!\!\!\!\!\!\!\!\!\!\!\!\!\!\!\!\!\!\!\!\!\!\!\!\!\!\!\!\!\!\!\!\!\! 
+ {m^2 \over 2 k_\mu q^3} 
\left[4 \mu \epsilon_q - (\epsilon_q - \omega)^2 \right]
\ln \left({\epsilon_q + q v_\mu - \omega 
  \over \epsilon_q + q v_\mu - \omega} \right) \Bigg\},
\end{eqnarray}
where $\epsilon_q = q^2 /2m$, $\mu$ is the chemical potential and $\mu
= k_\mu^2 /2m = m v_\mu^2 /2$, and the frequency $\omega$ can be any
complex number.

Finite temperature polarizability can be easily obtained from those at
zero temperature using the following identity:
\begin{equation}
\label{eq:finiteTP}
\Pi (q, \omega, \mu; T) = \int\limits_0^\infty d \mu' 
{\Pi_0(q, \omega, \mu') \over 4 T \cosh^2 ({\mu' - \mu \over 2T}) }.
\end{equation}
We find Eq.~(\ref{eq:finiteTP}) to be the most convenient numerical
method for obtaining the finite-$T$ polarizability.


\subsection{Dimensionless parameters}
\label{sec:form:para}

Our 2D and 3D electron system can be characterized by two parameters,
namely density ($n$) and temperature ($T$). This immediately leads to
two dimensionless parameters $r_s$ and $T/T_F$ characterizing the
system with $r_s$ being the effective zero-temperature interaction
strength and $T/T_F$ being the effective temperature (note that they
are {\em not} independent). The definition of $r_s$ is the following.
In 2DES $r_s$ is defined such that
\begin{eqnarray}
\label{eq:2Drs}
\pi r_s^2 a_B^{2} n &=& 1,\\
k_F r_s a_B &=& \sqrt{2},
\end{eqnarray}
where $n$ is the 2D electron density, $k_F$ is the Fermi momentum, and
$a_B = (m e^2)^{-1}$ is the Bohr radius. In 3DES $r_s$ is defined such
that
\begin{eqnarray}
\label{eq:3Drs}
4 \pi n a_B^3 r_s^3 /3 &=& 1,  \\
k_F r_s a_B &=& (9 \pi /4)^{1/3}.
\end{eqnarray}
The Fermi temperature $T_F \equiv E_F \equiv k_F^2 /(2m)$, which goes
as $T_F \propto r_s^{-2}$ in both 2D and 3D.


\section{Numerical methods in $m^*$ calculations in RPA}
\label{sec:method}

In this section we explain in detail our numerical approach for the
effective mass calculation within RPA. In carrying out the
integrations of self-energy in Eq.~(\ref{eq:Sig_gen}) in order to
obtain the effective mass, we use three different techniques, namely
frequency sum, frequency integration, and plasmon-pole approximation
(PPA). The first two techniques are equivalent, and we explain them in
detail in this section. PPA is a further approximation of RPA, which
has been extensively used in the
literature~\cite{lundqvist,dassarma,vinter}. We discuss the PPA in
section~\ref{sec:PPA}. Since there is no existing literature on the
finite temperature effective mass or self-energy calculation to check
our numerical results, it is crucial for us to use these different
techniques to ensure the correctness of our numerical calculations. We
mention here that our frequency sum results and frequency integration
results agree well with each other. The frequency integration result
is numerically relatively more noisy and therefore in this paper we
will only show the frequency sum results. We also check our numerical
results against the already known results at $T=0$ and against
analytical calculations in the $T/T_F, r_s \to 0 $ limit.

\subsection{Frequency integration technique}
\label{sec:method:int}

Eq.~(\ref{eq:Sig_gen}) gives the general formula for the RPA
self-energy at real frequencies. It can also be written in a more
succinct way as

\begin{eqnarray}
\label{eq:Eint}
\Sigma ({\bf k}, \omega) &=& 
  - \int \frac{d^d q}{(2 \pi)^d} 
  v_0(q) n_F ( \xi_{{\bf q} - {\bf k}} ) \nonumber \\
  && -\int \frac{d^d q}{(2 \pi)^d} \int \frac{d \varepsilon}{2 \pi}
   \frac{ 2 v_0(q) \mbox{Im} \epsilon^{-1} (q, \varepsilon )} {
     \varepsilon + \omega + i 
     \eta - \xi_{{\bf q} - {\bf k}} } \nonumber \\
  &&~~~~\cdot \left[ n_F ( \xi_{{\bf q} - {\bf k}} ) + n_B ( \varepsilon )
  \right], 
\end{eqnarray}
where $n_F(x) = 1/(\exp(x/T) + 1)$ is the Fermi function and $n_B(x) =
1/(\exp(x/T) - 1)$ the Bose function. This method of calculating the
self-energy involves integration over real frequencies, and therefore
we call it the frequency integration method. It is also known as the
spectral or the Lehmann representation of the self-energy. The
derivation of Eq.~(\ref{eq:Eint}) from Eq.~(\ref{eq:Sig_gen}) is given
in the appendix.

The self-energy of Eq.~(\ref{eq:Eint}) is composed of two parts: the
exchange part and the correlation part. The (frequency independent)
exchange part is also known as the Hartree-Fock self-energy, and its
contribution to the effective mass at $T=0$ is singular in both 2D and
3D. Not surprisingly, this singularity is cancelled out by
contributions from the correlation part of the self-energy. Effective
mass is derived from the self-energy through Eq.~(\ref{eq:onshellm}),
and we therefore need to obtain the real part of Eq.~(\ref{eq:Eint})
by putting $i \eta$ to be $0$ and regarding the frequency integration
as a principal value integration. It is easy to derive from
Eq.~(\ref{eq:Eint}) that the imaginary part of the self-energy can be
written as
\begin{eqnarray}
\label{eq:Eimag}
\mbox{Im} \Sigma({\bf k}, \omega) &=& - \int \frac{d^d q}{(2 \pi)^d} v_0(q)
\mbox{Im}\epsilon^{-1}({\bf q}, \xi_{{\bf q} - {\bf k}} - \omega))
\nonumber \\
&&~~~~~\cdot \left[n_B(\xi_{{\bf q} - {\bf k}} - \omega)
    +n_F(\xi_{{\bf q} - {\bf k}}) \right].
\end{eqnarray}
The ${\rm Im} \Sigma$ is not needed in the effective mass calculation
since $m^*$ is a Fermi surface property. But it is important to have
some idea of the magnitude of ${\rm Im} \Sigma$ in order to ensure
that quasi-particles are well defined at finite $T$.

Numerically carrying out the integration in Eq.~(\ref{eq:Eint}) is
non-trivial: for each momentum $\bf q$ and frequency $\omega$, a three
dimensional integration is required to obtain $\Sigma({\bf q},
\omega)$, and what makes the problem even more difficult is that the
$\mbox{Im} \epsilon^{-1}({\bf q}, \omega) $ term in the integrand is
highly non-monotonic. A careful examination of the dynamical
dielectric function tells us that at $T=0$, $\mbox{Im}
\epsilon^{-1}({\bf q}, \omega)$ contains delta-functions at plasmon
excitation frequencies, and at finite temperatures these
delta-functions broaden into sharp peaks. Integration over these
sharps peaks requires special care. For each ${\bf q}$, the position
(i.e., frequency) of the sharp peaks can be determined by solving
$\mbox{Re} [\epsilon({\bf q}, \omega)] = 0$, and their weight can be
determined from $\mbox{Re} [\epsilon({\bf q}, 0)]$ using the
Kramers-Kr\"{o}nig relations.

One advantage of the frequency integration method is that in
Eq.~(\ref{eq:Eint}) we can directly put $T=0$ to obtain the zero
temperature result, in contrast to the frequency sum method which we
will describe in detail below.


\subsection{Frequency sum technique}
\label{sec:method:sum}

Due to the great numerical difficulty in carrying out the frequency
integration method introduced above (because it involves integration
over highly non-monotonic or singular functions), it is advisable to
seek alternatives. At zero temperature, previous works in calculating
self-energy and related quantities often transform the real frequency
integration into integrations over imaginary frequencies using the
analytic properties of the dielectric function. The purpose of this
contour distortion is to avoid singularities along the real axis. At
finite temperature, a similar approach can be adopted. At finite
temperature, what is different from the zero temperature case is that
we transform the integration into an imaginary frequency summation (or
Matsubara frequency summation). Hu {\it at al}.~\cite{hu} showed in
detail how to perform such a transformation from the real-frequency
integration to an imaginary frequency summation. Following the
technique of contour distortion introduced in Ref.~\cite{hu} we can
write the RPA self-energy as a sum of the Matsubara frequency along
the imaginary axis:
\begin{eqnarray}
  \label{eq:Esum}
  \Sigma({\bf k}, \omega) =&-& \int \frac{d^d q}{(2 \pi)^d}
  v_0(q) n_F(\xi_{{\bf q} - {\bf k}}) \nonumber \\
  &-& \int \frac{d^d q}{(2 \pi)^d} v_0(q) \left[
    \frac{1}{\epsilon(q, \xi_{{\bf q} - {\bf k}} - \omega)} - 1 \right]
  \nonumber \\
  &&~~~~~\cdot \left[n_B(\xi_{{\bf q} - {\bf k}} - \omega)
    +n_F(\xi_{{\bf q} - {\bf k}}) \right] \nonumber \\
  &-& \int \frac{d^d q}{(2 \pi)^d} T \sum_{\omega_n}
  v_0(q) \left[ \frac{1}{\epsilon(q, i \omega_n)} - 1 \right]
  \nonumber \\
  && ~~~~ \cdot \frac{1}{i \omega_n - (\xi_{{\bf q} - {\bf k}} -
  \omega)},
\end{eqnarray}
where the frequency sum is over even Matsubara frequencies $i \omega_n
= i 2 n \pi T$ with $n$ integers. The above expression contains three
parts, namely the exchange part, the residue part and the line part
from top to bottom in Eq.~(\ref{eq:Esum}). The proof of the
equivalence between Eq.~(\ref{eq:Esum}) and Eq.~(\ref{eq:Eint}) is
provided below.

\begin{figure}[htbp]
\centering \includegraphics[width=2in]{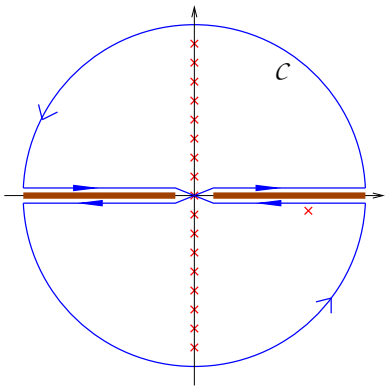}
\caption{Contour of integration for the derivation of self-energy
  formula for the frequency sum method. The thick lines on real axis
  denotes the branch cut for $\epsilon^{-1}({\bf q}, \omega)$. The
  crosses mark the poles due to the integrand; the ones on the
  imaginary axis are due to $n_B(\omega)$, and the isolated pole is
  due to the denominator.}
\label{fig2}
\end{figure}

Since the exchange part exists in both Eq.~(\ref{eq:Esum}) and
Eq.~(\ref{eq:Eint}), we only need to consider the correlation part of
the self-energy
\begin{eqnarray}
\label{eq:Esum1}
\Sigma^{\mbox{cor}} ({\bf k}, \omega) &=& 
   -\int \frac{d^d q}{(2 \pi)^d} \int \frac{d \nu}{2 \pi}
   \frac{ 2 v_0(q) \mbox{Im} \epsilon^{-1} (q, \nu )} { \nu + \omega + i
     \eta - \xi_{{\bf q} - {\bf k}} } \nonumber \\
  &&~~~~\cdot \left[ n_F ( \xi_{{\bf q} - {\bf k}} ) + n_B ( \nu )
  \right].
\end{eqnarray}
We choose the contour as in Fig.~\ref{fig2}. It is easy to see that
the integration over real axis can be transformed into integration
over contour $\cal C$, so that we have
\begin{eqnarray}
\label{eq:Esum2}
\Sigma^{\mbox{cor}} ({\bf k}, \omega) &=& 
   -\int \frac{d^d q}{(2 \pi)^d} \oint_{\cal C} \frac{d \nu}{2 \pi i}
   \frac{ v_0(q) (\epsilon^{-1} (q, \nu )-1)} { \nu + \omega + i
     \eta - \xi_{{\bf q} - {\bf k}} } \nonumber \\
  &&~~~~\cdot \left[ n_F ( \xi_{{\bf q} - {\bf k}} ) + n_B ( \nu )
  \right].
\end{eqnarray}
This is because $\epsilon({\bf q}, \omega -i \eta)= \epsilon^*({\bf
  q}, \omega+i \eta)$, and therefore the integration of the real part
of the integrand right above the real axis in the positive direction
and right below the real axis in the negative direction cancel each
other, and the corresponding integration of the imaginary part on
these two lines are equal to each other. The $-1$ after
$\epsilon^{-1}({\bf q}, \omega)$ is inserted to make the integration
on the arc part of contour $\cal C$ vanish as the radius of the
contour approaches infinity. Now we are left to evaluate the residues
within contour $\cal C$, the positions of which are denoted by crosses
in Fig.~\ref{fig2}. Note that the analytic property of the dielectric
function $\epsilon^{-1}({\bf q}, \omega)$ is very important in this
approach. The transformation requires that $\epsilon^{-1}({\bf q},
\omega)$ is analytic in the upper and the lower half of the complex
plane, which is true for electron gas systems. The single residue at
$\xi_{{\bf q} - {\bf k}} - \omega - i \eta$ right below the real axis
produced by the denominator of the integrand produces the residue part
of the self-energy. This part can be easily derived as
\begin{eqnarray} 
\label{eq:Eres}
\Sigma^{\mbox{res}} &=& - 
  \int \frac{d^d q}{(2 \pi)^d} v_0(q) \left[
    \frac{1}{\epsilon(q, \xi_{{\bf q} - {\bf k}} - \omega)} - 1
  \right] \nonumber \\
  &&~~~~~\cdot \left[ n_B(\xi_{{\bf q} - {\bf k}} - \omega)
    +n_F(\xi_{{\bf q} - {\bf k}}) \right].
\end{eqnarray}
The residues at $\omega_n = 2 n \pi T$ on the imaginary axis (the
third term in Eq.~(\ref{eq:Esum})), which are produced by the Bose
function $n_B(\nu)$, lead to the line part of the self-energy. This
part can be written as
\begin{eqnarray} 
\label{eq:Eline}
\Sigma^{\mbox{line}} &=& - 
  \int \frac{d^d q}{(2 \pi)^d} T \sum_{\omega_n}
  v_0(q) \left[ \frac{1}{\epsilon(q, i \omega_n)} - 1 \right]
  \nonumber \\
  && ~~~~ \cdot \frac{1}{i \omega_n - (\xi_{{\bf q} - {\bf k}} -
  \omega)}.
\end{eqnarray}
From Eq.~(\ref{eq:Eres}) and Eq.~(\ref{eq:Eline}) we have
$\Sigma^{\mbox{cor}} = \Sigma^{\mbox{res}} + \Sigma^{\mbox{line}}$,
and we thus obtain Eq.~(\ref{eq:Esum}).

The frequency sum method proves to be a far more efficient numerical
technique for calculating the self-energy than the frequency
integration method due to the absence of the strong non-monotonicity
and singularity in the real frequency dependence of the integrand. One
thing to notice is that at high temperatures, higher Matsubara
frequency terms can be neglected because $(\epsilon^{-1}({\bf q}, i
\omega_n) - 1) \to 0$ when $\omega_n \to \infty$, while at low
temperatures a large number of Matsubara terms have to be kept in the
sum in order to ensure accuracy. At zero temperature, the frequency
sum turns into an integration over imaginary frequencies, and we have
\begin{eqnarray}
\label{eq:Esum0}
\Sigma ({\bf k}, \omega) &=& - \int_{{\cal R}_1} 
\frac{d^d q}{(2 \pi)^d} v_0(q) \nonumber \\
&&+ \int_{{\cal R}_2} \frac{d^d q}{(2 \pi)^d} \frac{v_0(q)}
{\epsilon({\bf q}, \xi_{{\bf q} - {\bf k}} - \omega)} \nonumber \\
&&-\int \frac{d^d q}{(2 \pi)^d} \int \frac{d \nu}{2 \pi}
\left[ \frac{1}{\epsilon({\bf q}, i \nu)} - 1 \right] \nonumber \\
&&~~~~~~\cdot \frac{\omega - \xi_{{\bf q} - {\bf k}}}
{\nu^2  + (\omega -\xi_{{\bf q} - {\bf k}})^2 },
\end{eqnarray}
where the integration region ${\cal R}_1$ denotes the region where
$|{\bf k} - {\bf q}| < k $, and ${\cal R}_2$ denotes the integration
region where$|{\bf k} - {\bf q}|$ is in between $k$ and $k_F$. This
explicit formula for self-energy is exactly what previous works (see,
e.g. Ref.~\cite{ting}) used to calculate the zero temperature RPA
self-energy.

It is obvious that the frequency independent exchange part of the
self-energy is real. By noticing that $\epsilon({\bf q}, -\omega_n) =
\epsilon^*({\bf q}, \omega_n)$, we can see that the line part of the
self-energy is real as well. Thus the only contribution to the
imaginary part of the self-energy comes from the residue part, which
gives the same result as Eq.~(\ref{eq:Eimag}) in the frequency
integration method.


\section{RPA Results for $m^*(r_s, T/T_F)$}
\label{sec:results}

In this section we present our numerical results for effective mass in
2D and 3D electron systems within RPA. We first present in
section~\ref{sec:result:zeroT} results for the zero temperature
effective mass to compare with earlier works. Our finite temperature
results for 2DES are presented in section~\ref{sec:result:2D} and
those for 3DES in section~\ref{sec:result:3D}. In
section~\ref{sec:result:beyond} we present results for a model bare
potential where the Coulomb interaction is cut-off by a finite length
so that the bare interaction is short-ranged. We do this in order to
investigate the model dependence of our results.

\subsection{Zero temperature effective mass}
\label{sec:result:zeroT}

\begin{figure}[htbp]
\centering \includegraphics[width=2.5in]{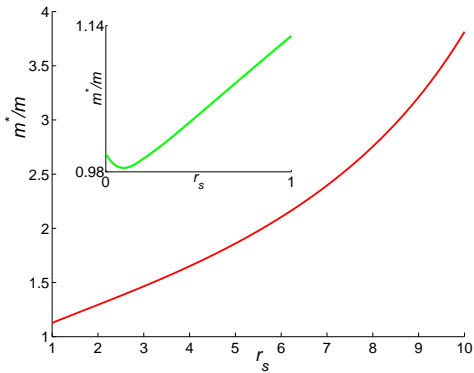}
  \caption{Calculated $T \sim 0$ effective mass as a function of $r_s$
    in a 2DES. Inset: the result in low $r_s$ region.}
\label{fig3}
\end{figure}

\begin{figure}[htbp]
\centering \includegraphics[width=2.5in]{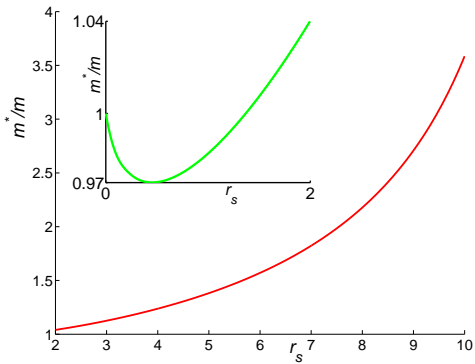}
  \caption{Calculated $T \sim 0$ effective mass as a function of $r_s$
    in a 3DES. Inset: the result in low $r_s$ region.}
\label{fig4}
\end{figure}

We first present our extreme low temperature results ($T/T_F \approx
10^{-4}$) in Fig.~\ref{fig3} and Fig.~\ref{fig4}, to be compared with
the existing $T=0$ results
\cite{rice,gellmann,ting,vinter,jalabert,marmorkos,book}. We calculate
$m^*(r_s)$ in the $r_s = 0 - 10$ range, showing that the effective
mass renormalization could be as large as 4.5 for dilute ($r_s \sim
10$) 2DES and 3 for ($r_s \sim 10$) 3DES. We emphasize that the
results presented in Fig.~\ref{fig3} and Fig.~\ref{fig4} are entirely
based on the $T \to 0$ limit of our finite temperature theory. They
are in {\it quantitative} agreement with the existing $T = 0$ 2D RPA
effective mass calculations \cite{ting} (which are restricted to the
$r_s < 5$ regime) and are consistent with the existing $T = 0$ 3D
effective mass calculations at low $r_s$~\cite{rice}. This serves as a
stringent check on our numerical approaches.

It is clear from Figs.~\ref{fig3} and \ref{fig4} that both our 2D and
3D results show the non-monotonic dependence of $m^*(r_s)$ on $r_s$ in
the high-density regime (i.e. in the $r_s \ll 1$ regime). This
nonmonitonic low-$r_s$ behavior for $m^*(r_s)$ at $T=0$ has been
reported in the earlier literature~\cite{rice,ting}.

We emphasize that the numerical results given in Figs.~\ref{fig3} and
\ref{fig4} are obtained by putting $T/T_F \approx 10^{-4}$ in our
finite-temperature formalism.


\subsection{Finite temperature effective mass in 2DES}
\label{sec:result:2D}

\begin{figure}[htbp]
\centering \includegraphics[width=2.5in]{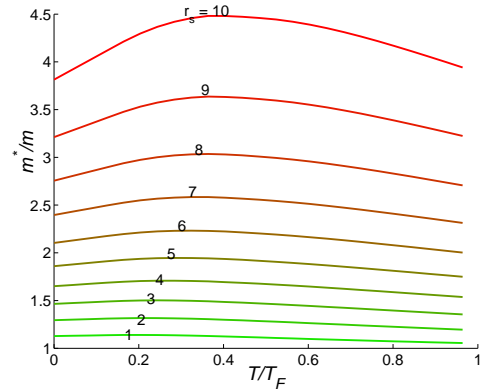}
  \caption{Calculated 2D effective mass as a function of $T/T_F$ for
    different $r_s$: $r_s = 10 \to 1$ from top to bottom; Inset: $r_s
    = 5 - 1 $ from top to bottom. Note that $T_F \propto r_s^{-2}$,
    making the absolute temperature scale lower for higher $r_s$
    values.}
  \label{fig5}
\end{figure}

\begin{figure}[htbp]
\centering \includegraphics[width=2.5in]{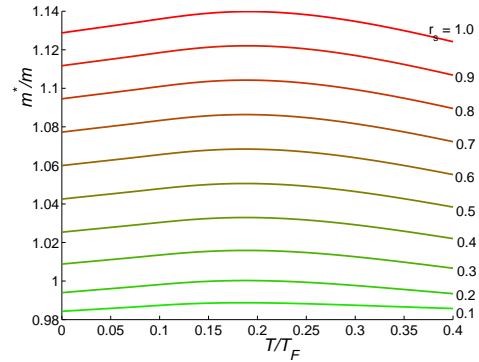}
  \caption{Calculated 2D effective mass as a function of $T/T_F$ for
    low $r_s$ values: $r_s = 1.0 \to 0.1$ from top to bottom.}
  \label{fig6}
\end{figure}

In Fig.~\ref{fig5} and Fig.~\ref{fig6} we show our calculated 2D
$m^*(T)$ as a function of $T/T_F$ for different values of the 2D
interaction parameter $r_s$ ($=0.1 - 10$). In the low temperature
region the effective mass first rises to some maximum, and then
decreases as temperature increases. This nonmonotonic trend is
systematic, and the value of $T/T_F$ where the effective mass reaches
the maximum increases with increasing $r_s$. The initial increase of
$m^*(T)$ is almost linear in $T/T_F$ as $T \to 0$, and the slope
$\frac{d(m^*/m)}{d(T/T_F)}$ is almost independent of $r_s$ for very
small $r_s$ ($<1$) (which is shown in Fig.~\ref{fig6}), but increases
with $r_s$ for larger $r_s$ values. It is important to notice that
this non-monotonic temperature dependence of $m^*(T)$ with a maximum
around $T/T_F \lesssim 1$ persists all the way to $r_s \to 0$, which
suggests that it is not an artifact of our approximation scheme since
RPA become {\it exact} as $r_s \to 0$.

\begin{figure}[htbp]
\centering \includegraphics[width=2.5in]{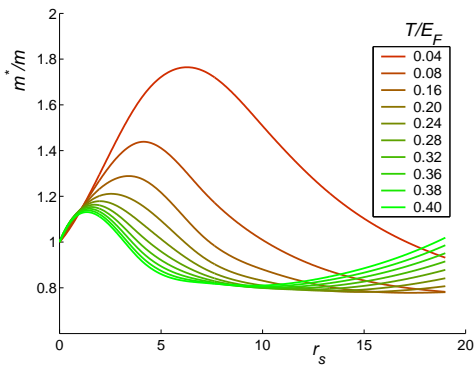}
  \caption{Calculated 2D effective mass as a function of $r_s$ at
    fixed value of temperatures. T is in the unit of $T_F$ at $r_s =
    1$.}
  \label{fig7}
\end{figure}

\begin{figure}[htbp]
\centering \includegraphics[width=2.5in]{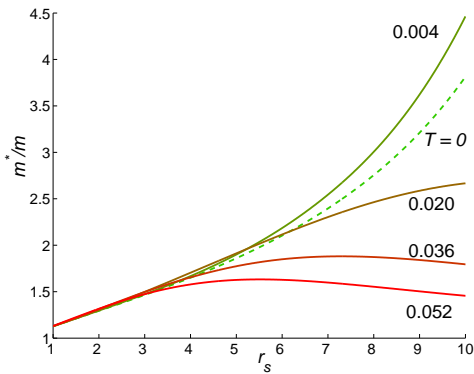}
  \caption{Calculated 2D effective mass as a function of $r_s$ at
    fixed value of temperatures. T is in the unit of $T_F$ at $r_s =
    1$. This plot is similar to Fig.~\ref{fig7} but concentrating on
    the low temperature region.}
  \label{fig8}
\end{figure}

In Fig.~\ref{fig7} and Fig.~\ref{fig8} we show the dependence of the
effective mass renormalization as a function of the interaction
parameter $r_s$ for a few values of {\it fixed} temperature (rather
than fixed $T/T_F$, remembering that $T_F \propto r_s^{-2}$ since $T_F
\propto n$ and $r_s \propto n^{-1/2}$). Fig.~\ref{fig7} shows the
effective mass for high $T$ and large $r_s$ values while
Fig.~\ref{fig8} concentrates on the low $T$ region. The calculated
$m^*(r_s)$ for fixed $T$ values are quite striking: For low fixed
values of $T$, $m^*/m$ initially increases with $r_s$ even faster than
the corresponding $T=0$ result, eventually decreasing with $r_s$ at
large enough values (where the corresponding $T/T_F$ values become
large enough). This nonmonotonic behavior of $m^*(r_s)$ as a function
of $r_s$ for fixed temperatures showing a temperature-dependent
maximum (with the value of $r_s$ at which the $m^*$ peak occurs
decreasing with increasing $T$ as in Fig.~\ref{fig7}) is complementary
to the nonmonotonicity of $m^*(T)$ in Fig.~\ref{fig5} as a function of
$T/T_F$ (at fixed $r_s$) and arises from the relationship between the
dimensionless variables $T/T_F$ ($\propto r_s^{-2}$) and $r_s$
($\propto T_F^{-1/2}$) due to their dependence on the carrier density
(i.e. $T_F \propto n$ and $r_s \propto n^{-1/2}$). At large $r_s$ and
high temperature, Fig.~\ref{fig7} shows that the effective mass
increases from below unity with increasing $r_s$. This is the region
where the exchange part of the self-energy dominates, and it can be
easily shown that the exchange self-energy produces this peculiar
effect on the $r_s$ dependence of $m^*(r_s)$ at fixed high $T$ values.
Since this region is hardly accessible by experiments, and moreover
the quasiparticles may not even be well-defined at such high $T/T_F$
values, we do not further discuss the physics related to this region.

One immediate consequence of our results shown in Figs.~\ref{fig5} and
\ref{fig7} is that $m^* (T/T_F, r_s) \equiv m^*(T, n)$ in 2DES could
show a strong enhancement at low (but finite) temperatures and low
electron densities (large $r_s$). Comparing with the actual system
parameters for 2D electrons in Si inversion layers and GaAs
heterostructures (and taking into account the quasi-2D form factor
effects~\cite{ando} neglected in our strictly 2D calculation) we find
that, consistent with recent experimental findings~\cite{exp}, our
theoretical calculations predict (according to Figs.~\ref{fig5} and
\ref{fig7} as modified by subband form factors) $m^*/m$ to be enhanced
by a factor of $2-4$ for the experimental densities and temperatures
used in recent measurements~\cite{exp}. Due to the approximate (i.e.
RPA) nature of our theory we do not further pursue the comparison with
experimental data in this paper since the main goal of this paper is
to discuss the temperature dependence of $m^*(r_s, T/T_F)$ which has
not yet been reported in the literature. A direct experimental
observation of an increasing $m^*(T)$ at low temperatures in 2DES will
be a striking confirmation of our theory.


\subsection{Finite temperature effective mass in 3DES}
\label{sec:result:3D}

\begin{figure}[htbp]
\centering \includegraphics[width=2.5in]{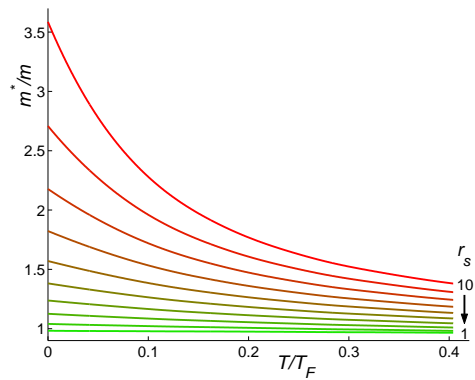}
  \caption{Calculated 3D effective mass as a function of $T/T_F$ for
    different $r_s$: $r_s = 10 \to 1$ from top to bottom.}
  \label{fig9}
\end{figure}

\begin{figure}[htbp]
\centering \includegraphics[width=2.5in]{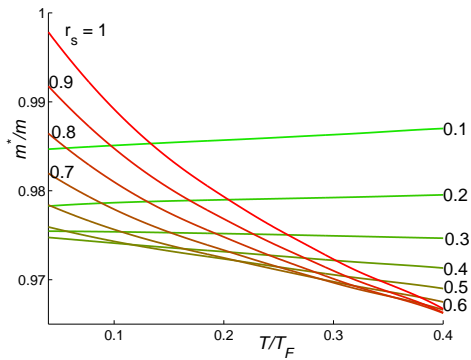}
  \caption{Calculated 3D effective mass as a function of $T/T_F$ for
    low $r_s$ values: $r_s = 1.0 \to 0.1$.}
  \label{fig10}
\end{figure}

In Fig.~\ref{fig9} and Fig.~\ref{fig10} we show our calculated 3D
$m^*(T)$ as a function of $T/T_F$ for different $r_s$ values. In
Fig.~\ref{fig9} $r_s$ varies from $1$ to $10$ while in
Fig.~\ref{fig10} $r_s$ is from $0.1$ to $1$. The 3D temperature
dependence of the effective mass shows very different characteristics
from that of 2D. Fig.~\ref{fig9} shows that for $r_s > 1$ the
effective mass decreases monotonically with increasing $T$ at low
temperatures. However for $r_s << 1$, as shown in Fig.~\ref{fig10},
the effective mass increases with increasing $T$ in the temperature
region we are considering. We therefore conclude that in 3DES the sign
of the slope $\frac{d(m^*/m)}{d(T/T_F)|_{T=0}}$ is non-universal,
which differs from that of 2DES where the above mentioned slope is
always positive for all $r_s$. Another interesting feature is that the
sign of $\frac{d(m^*/m)}{d(T/T_F)}|_{T=0}$ matches the sign of
$-\frac{d(m^*/m)}{d(r_s)}|_{T=0}$ very well. In particular, $m^*(T)$
decreases with increasing $T$ (at low $T$) in the ``larger''
$r_s$-regime where the corresponding $T=0$ $m^*(r_s)$ shows an
increasing mass with increasing $r_s$. Similarly, $m^*(T)$ increases
(at low $T$) with increasing $T$ in the $r_s$-regime where the
corresponding $m^*(r_s; T=0)$ shows decreasing $m^*$ with increasing
$r_s$.

\subsection{Model short-range bare interaction}
\label{sec:result:beyond}

So far in all of our calculations we have been using the realistic
long-ranged Coulomb interaction for the bare potential as in
Eq.~(\ref{eq:bareV}). A question naturally arises: how is the
temperature dependence that we find in our calculations related to the
long range nature of the interaction between electrons? Therefore we
also calculate the effective mass in 2DES and 3DES using simple
(parameterized) finite-range interaction model
\begin{eqnarray}
\label{eq:newV}
v^{\mbox{2D}}(q)  &=& {2 \pi e^2 \over q + a}, \nonumber \\
v^{\mbox{3D}}(q)  &=& {4 \pi e^2 \over q^2 + a^2}.
\end{eqnarray}
where $a$ is the cut-off wavevector which eliminates the long
wavelength Coulomb divergence.

Our numerical calculation shows that as $a/k_F \to 0$, we recover the
$m^*(T)$ behavior of the bare Coulomb interaction results in both 2D
and 3D. As $a/k_F$ increases, the mass renormalization in both 2D and
3D is suppressed, but all the qualitative features of the temperature
dependence persist. In 2DES, as $a/k_F$ increases, the temperature
where the effective mass reaches the maximum decreases, and the
effective mass enhancement (from the $T=0$ value to the maximum)
decreases, but the linear-$T$ dependence at low $T$ and the
non-monotonic trend remain unchanged. In 3DES, as $a/k_F$ increases,
the $r_s$ region where $\frac{d(m^*/m)}{d(r_s)}|_{T=0} < 0$ shrinks,
but the consistency between the sign of
$-\frac{d(m^*/m)}{d(r_s)}|_{T=0}$ and the sign of
$\frac{d(m^*/m)}{d(T/T_F)}|_{T=0}$ remains.

From these result we conclude that the qualitative features of the
temperature dependence are model independent and {\em not} peculiar to
the bare interaction being Coulombic. This conclusion is further
reinforced by the recent report of a linearly $T$ dependent electronic
specific heat in a short-range interaction model~\cite{chubukov}. It
may be worth while, however, to note that RPA is specific to the
long-range Coulomb interaction in giving an exact result in the
high-density $r_s \to 0$ limit, and there is nothing special about RPA
in the case of short-range interaction.


\section{Plasmon-Pole approximation}
\label{sec:PPA}

We now apply a simple-to-use dynamical approximation to calculate
$m^*(T)$. The plasmon-pole approximation (PPA) has often been
used~\cite{vinter,dassarma,lundqvist} to obtain the electron
self-energy in the literature. It is a simple technique for carrying
out the frequency sum or integration in the RPA self-energy
calculation by using a spectral pole (i.e. a delta function) ansatz
for the dynamical dielectric function $\epsilon({\bf k}, \omega)$. In
other words, it is an approximation to the RPA. The PPA ansatz assumes
that
\begin{equation}
  \label{eq:ansatz}
  - 2 \mbox{Im}\frac{1}{\epsilon({\bf k}, {\omega})} =
  C_k (\delta(\omega - \bar{\omega}_k) - \delta(\omega + \bar{\omega}_k)),
\end{equation}
where the pole $\bar{\omega}_k$ and the spectral weight $C_k$ of the
PPA propagator in Eq.~(\ref{eq:ansatz}) are determined by using the
the Kramers-Kr\"{o}nig relation (i.e. causality)
\begin{equation}
  \label{eq:KK}
  \mbox{Re} \frac{1}{\epsilon(k, 0)} = 1 + \frac{2}{\pi} \int_0^{\infty}
  \frac{1}{\omega} d \omega \mbox{Im} \frac{1}{\epsilon(k, \omega)}
\end{equation}
and the $f$-sum rule (i.e. current conservation)
\begin{equation}
  \label{eq:sumrule}
  \int_0^{\infty} \omega d \omega \mbox{Im}
  \frac{1}{\epsilon(k, \omega)}
  = - \frac{\pi}{2} \omega_P^2(k).  
\end{equation}
Putting Eq.~(\ref{eq:ansatz}) in Eq.~(\ref{eq:KK}) and
Eq.~(\ref{eq:sumrule}) we have
\begin{eqnarray}
\label{eq:Ck}
C_k &=& \pi \omega_P(k) \sqrt{1 - \mbox{Re} \epsilon^{-1} (k, 0)}, \\
\label{eq:omegabar}
\bar{\omega}_k &=& \frac{\omega_P(k)}{\sqrt{1 - \mbox{Re} 
    \epsilon^{-1} (k, 0) }},
\end{eqnarray}
where $\omega_P(k)$ in Eqs.~(\ref{eq:sumrule}), (\ref{eq:Ck}) and
(\ref{eq:omegabar}) is the long wavelength plasmon frequency which is
defined as
\begin{equation}
\label{eq:omegaP}
\lim_{\omega \to \infty} \mbox{Re} [\epsilon(k, \omega)] 
= 1 - \frac{\omega_P^2(k)}{\omega^2}.
\end{equation} 
It is well-known that in 2DES
\begin{equation}
\label{eq:omegaP2D}
\omega_P^2(k) = \frac{2 \pi n e^2}{m} k,
\end{equation}
and in 3DES
\begin{equation}
\label{eq:omegaP3D}
\omega_P^2(k) = \frac{4 \pi n e^2}{m}. 
\end{equation}

We mention that $\bar{\omega}_k$ in Eq.~(\ref{eq:ansatz}) does {\it
  not} correspond to the real plasmon dispersion in the electron
liquid, but simulates the whole excitation spectra of the system
behaving as an effective plasmon at low momentum and as the
single-particle electron-hole excitation at large momentum, as
constrained by the Kramers-Kr\"{o}nig relation and the $f$-sum rule.
Details on the PPA are available in
literature~\cite{vinter,dassarma,lundqvist}, including its
finite-temperature generalization~\cite{dassarma}. The PPA, which is
known to give results close to the full RPA calculation of
self-energy, allows a trivial carrying out of the frequency sum in the
retarded self-energy function leading to:
\begin{eqnarray}
\label{eq:EPPA}
  \mbox{Re}\Sigma({\bf k}, \omega) =&-& \int \frac{d^2 q}{(2 \pi)^2}
  v_0(q) n_F(\xi_{{\bf q} - {\bf k}}) \nonumber \\
  &+& \int \frac{d^2 q}{(2 \pi)^2} v_0(q) C_q \Big[ 
  \frac{ n_B( \bar{\omega}_q ) + n_F ( \xi_{{\bf q} - {\bf k}} ) }
  {\bar{\omega}_q - ( \xi_{{\bf q} - {\bf k}} - \omega ) } \nonumber \\
&&~~~~~~~+ 
  \frac{ n_B( -\bar{\omega}_q ) + n_F ( \xi_{{\bf q} - {\bf k}} ) }
  {\bar{\omega}_q + ( \xi_{{\bf q} - {\bf k}}  - \omega) } \Big],
\end{eqnarray}
where $C_q$ and $\bar{\omega}_k$ only depend on $\epsilon(k, 0)$ at
finite temperatures, and are determined by Eq.~(\ref{eq:Ck}) and
Eq.~(\ref{eq:omegabar}). Obviously the PPA provides a great
simplification of the problem since the most numerically demanding
part of the calculation (the frequency sum or integration) is
trivially done. It should be noted, however, that although the PPA is
known to produce a reliable approximation to ${\rm Re} \Sigma$, it, by
definition, fails completely for ${\rm Im} \Sigma$.

\begin{figure}[htbp]
\centering \includegraphics[width=2.5in]{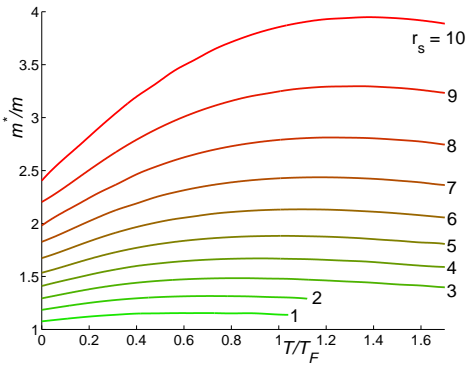}
  \caption{Calculated 2D PPA effective mass as a function of $T/T_F$ at
    fixed value of $r_s$.}
  \label{fig11}
\end{figure}

We present our PPA results for the 2D effective mass as a function of
$T/T_F$ at fixed $r_s$ values in Fig.~\ref{fig11}. One immediate
observation by comparing Fig.~\ref{fig5} and Fig.~\ref{fig11} is that
even though PPA provides a very good approximation for the self-energy
(indeed, our numerical results for PPA self-energy and RPA self-energy
match very well), it fails to provide accurate result for the
effective mass. The zero temperature effective mass generated by PPA
is almost half of that from RPA, and the temperatures where $m^*$
maximizes shift to higher $T$ values in the PPA compared with RPA. But
the qualitative behavior of $m^*(r_s, T/T_F)$ is similar in the PPA
and RPA for the 2DES as is clear by comparing Figs.~\ref{fig11} and
\ref{fig6}.

\begin{figure}[htbp]
\centering \includegraphics[width=2.5in]{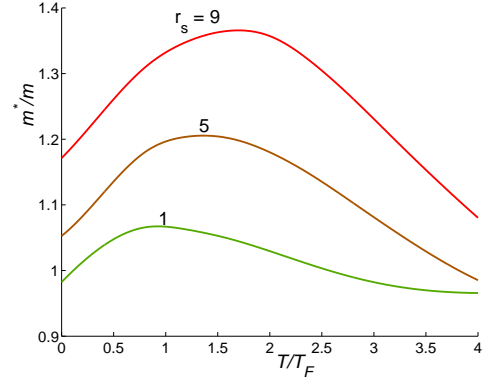}
  \caption{Calculated 3D PPA effective mass as a function of $T/T_F$ at
    fixed value of $r_s$.}
  \label{fig12}
\end{figure}

From our results of 3D PPA effective mass calculation presented in
Fig.~\ref{fig12} we can see that they are different from RPA results
even qualitatively. In fact, our RPA results for $m^*(r_s, T/T_F)$ are
similar in both 2D and 3D.

\section{Quasiparticle decay}
\label{sec:decay}

The quasiparticle decay rate (or the inverse lifetime) is
given~\cite{chaplik,gq} by the imaginary part of the self-energy. As
we have discussed in section~\ref{sec:method:sum}, the imaginary part
of the quasiparticle self-energy can be calculated from
Eq.~(\ref{eq:Esum}). It is also obvious that only the second term in
Eq.~(\ref{eq:Esum}) contributes to the imaginary self-energy: the
first term is obviously real, and the last term is also real because
$\epsilon({\bf q}, -\omega_n) = \epsilon^*({\bf q}, \omega_n)$. Thus
we have
\begin{eqnarray}
\label{eq:ImE}
  \mbox{Im} \Sigma({\bf k}, \omega) =
  &-& \int \frac{d^d q}{(2 \pi)^d} v_0(q) \mbox{Im}
    \frac{1}{\epsilon(q, \xi_{{\bf q} - {\bf k}})}
  \nonumber \\
  &&\cdot \left[n_B(\xi_{{\bf q} - {\bf k}} - \omega)
    +n_F(\xi_{{\bf q} - {\bf k}}) \right]. 
\end{eqnarray}

\begin{figure}[htbp]
\centering \includegraphics[width=2.5in]{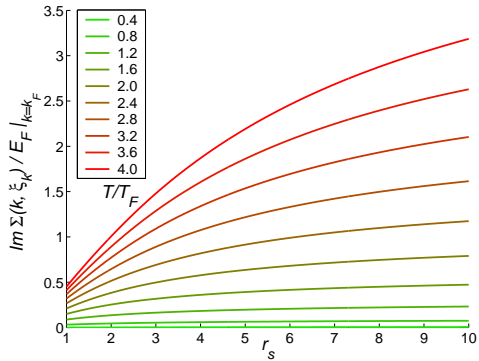}
  \caption{Calculated magnitude of the 2D RPA imaginary self-energy of
    quasiparticles on Fermi surface as a function of $r_s$ at
    different values of $T/T_F$.}
  \label{fig13}
\end{figure}

\begin{figure}[htbp]
\centering \includegraphics[width=2.5in]{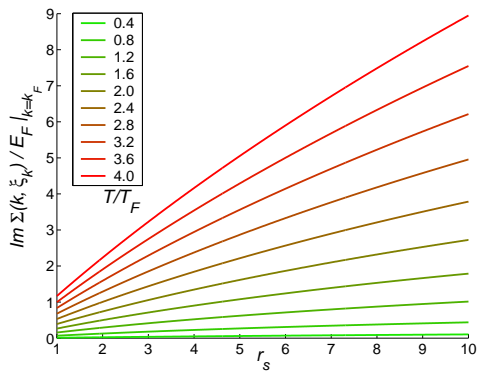}
  \caption{Calculated 3D RPA imaginary self-energy of quasiparticles on Fermi
    surface as a function of $r_s$ at different values of $T/T_F$.}
  \label{fig14}
\end{figure}

Fig.~\ref{fig13} and Fig.~\ref{fig14} show the calculated imaginary
self-energy on the Fermi surface in 2D and 3D respectively. The
quasiparticle decay (i.e. finite ${\rm Im} \Sigma$) here arises
entirely from having a finite temperature. The results show that the
magnitude of the imaginary self energy increases with increasing $r_s$
and $T/T_F$. It is obvious from Eq.~(\ref{eq:ImE}) that the imaginary
self-energy vanishes on the Fermi surface at $T = 0$ as it must since
the quasiparticles are perfectly well-defined at $T=0$ for $k=k_F$. As
$T$ increases, the magnitude of imaginary self-energy remains small
compared to the Fermi energy up to a certain temperature, and the
quasiparticles on the Fermi surface remain well-defined up to that
temperature. The important question is whether the finite temperature
quasiparticles are sufficiently well-defined for the interesting
behavior of $m^*(T)$ we discussed in section~\ref{sec:results} to be
experimentally observable. If the quasiparticles are ill-defined (i.e.
${\rm Im} \Sigma(k_F) > E_F$ in the temperature regime of interest)
then obviously all the interesting temperature dependence of $m^*(T)$
predicted by us is only of academic interest since the large
broadening will make it impossible to define quasiparticles, let alone
their effective mass. Examining the results of Figs.~\ref{fig13} and
\ref{fig14} compared with those presented in section~\ref{sec:results}
it is clear that there is a well-defined regime of $(r_s, T/T_F)$
values where $m^*(T/T_F)$ shows non-trivial temperature dependence
with the condition $E_F \gg \left| {\rm Im} \Sigma(k_F) \right|$
well-satisfied so that quasiparticles are well-defined. Although this
is not unexpected since $\left|{\rm Im} \Sigma(T) \right| \sim T^2$
for $T/T_F \ll 1$ whereas $m^*(T)/m \approx 1 + {\cal O}(T)$ in 2D, it
is nevertheless important to see that ${\rm Im} \Sigma$ remains small
in magnitude in the $(r_s, T/T_F)$ regime of interest.

Earlier theoretical work on the quasiparticle damping of 2D
interacting electron systems can be found in
Refs.~\cite{jalabert,marmorkos,chaplik,gq}.


\section{Discussion and conclusion}
\label{sec:con}

In this work we have obtained detailed results for the temperature
dependence of the quasiparticle effective mass, $m^*(r_s, T/T_F)$, at
arbitrary values of temperature and density in 2D and 3D electron
systems interacting via the long range Coulomb interaction. Our
central approximation is the RPA (i.e. the dynamically screened
Hatree-Fock self-energy approximation), which is the leading-order
one-loop self-energy calculation in a dynamically screened effective
interaction expansion. RPA is exact in the high-density ($r_s \to 0$)
limit at $T=0$, and is therefore a controlled nontrivial approximation
which is empirically known to work well for $r_s > 1$ (e.g. metals
with $r_s \sim 3 - 6$ and 2D semiconductors with $r_s \sim 1 - 10$).
We also calculate the finite-temperature imaginary self-energy (i.e.
the quasiparticle decay rate or broadening) to ensure that the
broadening remains small in the $(r_s, T/T_F)$ parameter regime of our
interest where $m^*(T)$ shows interesting temperature dependence.

As mentioned earlier in the paper, it is well-known that at $T=0$,
$m^*(r_s)$ can be exactly calculated (in both 2D and 3D) in the
asymptotic $r_s \to 0$ limit by systematically expanding the RPA
self-energy since ring diagrams (included in the RPA) are the most
divergent diagrams in the $r_s \to 0$ limit. Such a zero temperature
$r_s$-expansion of RPA gives the following formula for $m^*(r_s)$ in
both 2D and 3D:
\begin{equation}
\label{eq:m0T}
\left. {m^*(r_s) \over m} \right|_{r_s \to 0} = 1 
+ a r_s (b + \ln r_s) + {\cal O} (r_s^2),
\end{equation}
where $a$, $b$ are constants of order unity, What we find numerically
is that the leading temperature correction to this effective mass
formula is linear in $T/T_F$ in 2D and nonlinear in $T/T_F$ in 3D. In
this paper we have calculated $m^*(r_s, T/T_F)$ numerically for the
one-loop dynamically screened Hatree-Fock RPA self-energy theory for
arbitrary $r_s$ and $T/T_F$ finding nontrivial temperature dependence
of the effective mass at all densities.

Our most important result is the unexpected discovery of a strong
temperature-dependent quasiparticle effective mass $m^*(T)$ at low
temperatures in 2DES. Since the temperature scale for the temperature
dependence of $m^*(T)$ is the Fermi temperature which tends to be high
($\sim 10^4 K)$ in the 3D electron liquids (i.e. metals), our
temperature-dependent effective mass results for 3D systems are mostly
of theoretical interest since any actual $T$-dependence of $m^*(T)$ in
the $T/T_F \lesssim 10^{-4}$ regime will be miniscule. Our numerical
results for the calculated $m^*(T)$ in 2D systems are consistent with
a {\em linear} leading-order temperature correction for the 2D
quasiparticle effective mass: Results in Figs.~\ref{fig5} and
\ref{fig6} can be well fitted to the formula $m^*(T) \approx 1 +
A^{\rm 2D} (r_s) + B^{\rm 2D} (r_s) (T/T_F) + \cdots$ for small
$T/T_F$ where the slope $B^{\rm 2D} (r_s)$ seems to be a constant
independent of $r_s$ (i.e. density) at least in the high-density ($r_s
\ll 1$) limit; for $r_s > 1$ the slope $B^{\rm 2D} (r_s)$ has a weak
density dependence increasing somewhat with increasing $r_s$ (but our
approximation scheme, RPA, becomes less quantitatively reliable at
large $r_s$, therefore it is possible that the slope
$d(m^*/m)/d(T/T_F)$ is indeed independent of $r_s$ in the $T \to 0$
limit). In addition to this interesting (and unexpected) linear
leading-order temperature correction to the quasiparticle effective
mass, we also find $B^{\rm 2D} (>0)$ to be positive for all $r_s$,
indicting that in 2DES, the leading-order temperature correction to
the effective mass is positive. Thus, $m^*(T)$ increases with
increasing $T$ at first before eventually decreasing as $T/T_F$
increases substantially, leading to a maximum in $m^*(T)$ at some
intermediate temperature $T^*(r_s) \sim 0.5 T_F$ which is only weakly
density dependent (except of course through $T_F$ itself). All three
of these 2D findings (i.e. linear leading-order $T/T_F$ dependence of
$m^*$, increasing $m^*$ with $T/T_F$ at low temperatures, and the
nonmonotonic behavior with a maximum in $m^*(T/T_F)$ occurring at $T^*
\sim 0.5 T_F$) are surprising and unexpected. In principle, these
predictions can be experimentally tested since our calculations
presented in section~\ref{sec:decay} show that the quasiparticles
remain reasonably well-defined (i.e. the broadening, ${\rm Im}
\Sigma$, remains small) all the way to $T^*$ and perhaps even above
$T^*$. This is reasonable since the many-body correction to $m^*$ is
linear in $T/T_F$ whereas the broadening ${\rm Im} \Sigma \sim
(T/T_F)^2$, ensuring that for $T/T_F <1$, the quasiparticle effective
mass is a well-defined quantity. In contrast to the linear (with
positive slope) leading-order $T$-dependence we find for all $r_s$ in
our calculated 2D $m^*(T)$, our 3D results show non-universal
$m^*(r_s, T/T_F)$ behavior. In 3D, $m^*(T/T_F)$ increases with
increasing $T/T_F$ at low temperatures only for very high densities
(small $r_s$) -- for larger $r_s$ values $m^*(T)$ decreases
monotonically with increasing temperature (in sharp contrast to the
striking non-monotonicity in $m^*(T)$ in 2D) and this decrease is more
consistent with a nonlinear leading-order temperature dependence
(rather than a linear one as in 2D). Our best guess for our numerical
results shown in Fig.~\ref{fig9} and \ref{fig10} is the following
equation: $m^*/m \approx 1 + A^{\rm 3D} (r_s) + C^{\rm 3D} (T/T_F)^l
\ln (T_F/T) + \cdots$, where $l$ is a number of the order one (note
that numerically fixing the number $l$ is difficult and need much more
work), $C^{\rm 3D} > 0$ for $r_s < r_s^*$ and $C^{\rm 3D} < 0$ for
$r_s > r_s^*$ with $r_s^*$ being approximately the $r_s$ value where
$A^{\rm 3D}$ changes from being negative to positive.

\begin{figure}[htbp]
\centering \includegraphics[width=3in]{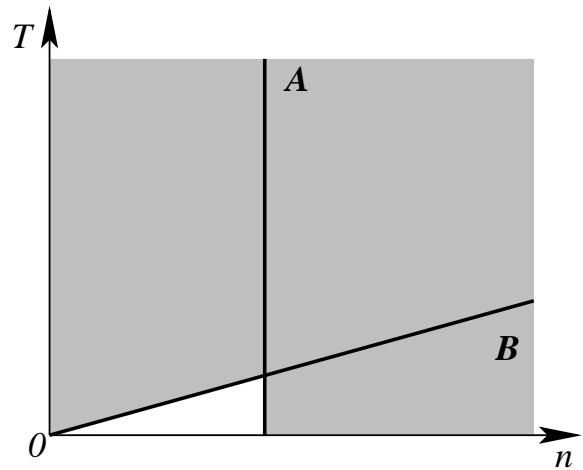}
  \caption{Schematic validity of RPA. The shaded area denotes the
    region where RPA is considered to be valid. Line A denotes a
    certain density value above which RPA is valid at $T=0$ (e.g. the
    vertical line A may correspond to the $r_s = 1$ condition so that
    for higher density, i.e. to the right to line A, RPA is valid even
    at $T=0$). Line B denotes the line of $T_F \propto n$. In the
    region above line B, RPA is again valid. Therefore for any fixed
    value of density $n$ (or equivalently fixed $r_s$), RPA is a
    better approximation with increasing $T$, whereas for fixed value
    of temperature, low density values (or large $r_s$)
    counter-intuitively makes RPA to valid again since RPA is valid
    for $T>T_F$.}
\label{fig15}
\end{figure}

We comment that our numerical results for $m^*(T)$ are consistent
(actually agree very well) with the very recent analytical
work~\cite{short,vd} on the temperature corrections to the effective
mass renormalization in 2D and 3D Fermi liquid. The analytical work is
necessarily restricted to the $r_s \to 0$ and $T/T_F \to 0$ limit
where the infinite series of ring diagrams to the electron self-energy
(depicted in Fig.~\ref{fig1}) provides an {\em exact} leading-order
asymptotic answer to the problem with the following result
\begin{eqnarray}
\label{eq:ana}
{m^*(r_s, T/T_F) \over m} = 1 &+& A(r_s) 
+ B(r_s) \left({T \over T_F} \right) \nonumber \\
&+& C(r_s) \left({T \over T_F} \right)^2 \ln \left({T \over T_F} \right) 
+ \cdots,
\end{eqnarray}
with $B(r_s) \equiv B^{\rm 2D}$, a constant, in 2D, and $B(r_s) \equiv
0$ in 3D. Our numerical results are consistent with this exact result,
but our numerical results apply also in the non-asymptotic region
where $T/T_F$ and $r_s$ are not necessarily small. In this
non-asymptotic regime (where $r_s$ is {\em not} small, actually $r_s$
may be large in 2D semiconductor systems) RPA is by no means an exact
theory, but we have recently argued~\cite{diverge} that RPA remains
extremely well-valid (if somewhat uncontrolled) even for $r_s \gg 1$.
We also emphasize a point in this context that seems not to have been
widely appreciated in the literature. The point is that RPA becomes a
progressively better approximation as $T/T_F$ increases at a fixed
$r_s$ (for any $r_s$), because the system is becoming more classical
in the $T/T_F \gg 1$ regime where RPA is again exact. Thus in the
$(r_s, T/T_F)$ parameter space (see Fig.~\ref{fig15}) RPA is exact as
$r_s \to 0$ (the high density limit) and as $T_F \to 0$ (the
high-temperature or equivalently the low density limit) or as $T \to
\infty$. Thus the regime of validity of RPA is greatly enhanced at
finite temperature, and in fact even at very large $r_s$ (i.e. very
low density) RPA becomes exact as $T$ is raised (because $T/T_F \gg 1$
limit is more easily achieved at low densities).

Finally, we comment on the anomalous (often referred to as
``nonanalytic''~\cite{chubukov}) nature of the temperature corrections
to the quasiparticle effective mass in 2D systems (but {\em not} in
3D) as manifested in the {\em linear} leading-order temperature
correction we find in interacting 2D electron systems.  This
particular feature is apparently generic in 2D and {\em not} due
merely to our using the long-range bare Coulomb interaction, because
in Ref.~\cite{chubukov} the same linear-$T$ correction is found in
calculations using a zero-range bare interaction although the sign of
the slope is negative in the zero-range interaction case. This kind of
(leading-order) linear temperature correction is quite common in 2D
electron systems due to the peculiar form of the 2D polarizability
with a $T=0$ cusp at $2k_F$. This leading-order linear-$T$ correction
is interesting because the naive expectation in a Fermi system (based
on the usual Sommerfeld expansion of the Fermi functions) is that the
leading-order correction in a ``normal'' situation should always be
${\cal O} (T/T_F)^2$ for all electronic properties. In 2D electron
systems it seems that the generic situation is ``anomalous'', i.e. the
leading-order temperature correction is ${\cal O} (T/T_F)$ rather than
the ``normal'' quadratic Fermi behavior expected on the basis of
analytic Sommerfeld expansion of Fermi functions. In 2D interacting
electron systems, therefore, all leading-order thermal corrections to
electronic properties are much stronger (by a factor of $T_F/T$, which
is a large number as $T \to 0$) than the quadratic Fermi gas behavior.
This anomalous nonanalyticity, which may have important consequences
for fermionic quantum critical phenomena, obviously has important
experimental implications since it is much easier to observe a linear
temperature correction than a quadratic one at low temperatures.

This work is supported by the US-ONR, the NSF, and the LPS.


\section*{Appendix}

Here we provide a proof of for equivalence between Eq.~(\ref{eq:Esum})
and Eq.~(\ref{eq:Eint}).

\begin{eqnarray}
\label{eq:Eint1}
&&\!\!\!\!\!\!\!\!\!\!
\Sigma ({\bf k}, \omega)= 
  -\int \frac{d^d q}{(2 \pi)^d} \int \frac{d \nu}{2 \pi}
  \Big\{ \nonumber \\
  &&~\mbox{Im} G_R({\bf k} - {\bf q}, \nu + \omega)
  D_R({\bf q}, \nu) 
  \tanh (\frac{\nu+\omega}{2 T} )
  \nonumber \\
  &&+G_R({\bf k} - {\bf q}, \nu + \omega)
  \mbox{Im}D_R({\bf q}, \nu) 
  \coth (\frac{\nu}{2 T}) \Big\} \nonumber
\end{eqnarray}
\begin{eqnarray}
  &=&-\int \frac{d^d q}{(2 \pi)^d} \int \frac{d \nu}{2 \pi} v_0(q)
  \Big\{ \nonumber \\
  &&- \pi \delta(\nu + \omega  - \xi_{{\bf q} - {\bf k}}) 
  \epsilon^{-1} ({\bf q}, \nu)  
    \tanh (\frac{\nu+\omega}{2 T} ) \nonumber \\
  &&+ \frac{1}{\nu + \omega + i \eta - \xi_{{\bf q} - {\bf k}}}
  \mbox{Im} \epsilon^{-1} ({\bf q}, \nu) 
  \coth ( \frac{\nu}{2 T} ) \Big\}.
\end{eqnarray}
Using Kramers-Kr\"{o}nig relations for $\epsilon^{-1}(q, \nu)$ in the
above equation, we have
\begin{eqnarray}
\label{eq:Eint2}
&&\!\!\!\!\!\!\!\!\!\!
\Sigma ({\bf k}, \omega) = 
  -\int \frac{d^d q}{(2 \pi)^d} \int \frac{d \nu}{2 \pi} v_0(q)
  \Big\{ \nonumber \\
  &&- \pi \delta(\nu + \omega  - \xi_{{\bf q} - {\bf k}})
  \nonumber \\
  && ~~~~\cdot \left[1 + \int \frac{d \nu'}{\pi} 
  \frac{ \mbox{Im} \epsilon^{-1} (q,\nu') }{\nu' - \nu - i \eta}
  \right] \tanh (\frac{\nu' +\omega}{2 T} ) \nonumber \\
  && + \frac{1}{\nu + \omega + i \eta - \xi_{{\bf q} - {\bf
        k}}}
  \mbox{Im} \epsilon^{-1} ({\bf q}, \nu) 
  \coth ( \frac{\nu}{2 T} ) \Big\} \nonumber 
\end{eqnarray}
\begin{eqnarray}
  &=& -\int \frac{d^d q}{(2 \pi)^d} v_0(q) \Big\{ \nonumber \\
  && -\frac{1}{2} \left[1 + \int \frac{d \nu}{\pi} \frac{ \mbox{Im}
    \epsilon^{-1} (q, \nu) }{\nu + \omega + i \eta
    - \xi_{{\bf q} - {\bf k}}} \right] 
    \tanh (\frac{\xi_{{\bf q} - {\bf k}}}{2 T} ) \nonumber \\
  && + \int \frac{d \nu}{2 \pi}
  \frac{1}{\nu + \omega + i \eta - \xi_{{\bf q} - {\bf k}}}
  \mbox{Im} \epsilon^{-1} ({\bf q}, \nu) 
  \coth ( \frac{\nu}{2 T} ) \Big\} \nonumber 
\end{eqnarray}
\begin{eqnarray}
  &=& \int \frac{d^d q}{(2 \pi)^d} v_0(q) \frac{1}{2}
  \tanh ( \frac{\xi_{{\bf q} - {\bf k}}}{2 T} ) \nonumber \\
  &&+\int \frac{d^d q}{(2 \pi)^d} \int \frac{d \nu}{2 \pi}
   \frac{ v_0(q) \mbox{Im} \epsilon^{-1} (q, \nu )} { \nu + \omega +
     i \eta - \xi_{{\bf q} - {\bf k}} } \nonumber \\
   &&~~~~~~~\cdot \left[ \tanh ( \frac{\xi_{{\bf q} - {\bf k}}}{2 T} ) 
   - \coth ( \frac{\nu}{2 T} ) \right] \nonumber 
\end{eqnarray}
\begin{eqnarray}
  &=& \mbox{Const} - \int \frac{d^d q}{(2 \pi)^d} 
  v_0(q) n_F ( \xi_{{\bf q} - {\bf k}} ) \nonumber \\
  &&~~~~~~~ -\int \frac{d^d q}{(2 \pi)^d} \int \frac{d \nu}{2 \pi}
   \frac{ 2 v_0(q) \mbox{Im} \epsilon^{-1} (q, \nu )} { \nu + \omega
     + i \eta - \xi_{{\bf q} - {\bf k}} } \nonumber \\
  &&~~~~~~~~~~~~~~~~~~~~~\cdot \left[ n_F ( \xi_{{\bf q} - {\bf k}} ) 
  + n_B ( \nu ) \right].
\end{eqnarray}


\bibliography{massprb}

\begin{thebibliography}{19}
\expandafter\ifx\csname natexlab\endcsname\relax\def\natexlab#1{#1}\fi
\expandafter\ifx\csname bibnamefont\endcsname\relax
  \def\bibnamefont#1{#1}\fi
\expandafter\ifx\csname bibfnamefont\endcsname\relax
  \def\bibfnamefont#1{#1}\fi
\expandafter\ifx\csname citenamefont\endcsname\relax
  \def\citenamefont#1{#1}\fi
\expandafter\ifx\csname url\endcsname\relax
  \def\url#1{\texttt{#1}}\fi
\expandafter\ifx\csname urlprefix\endcsname\relax\def\urlprefix{URL }\fi
\providecommand{\bibinfo}[2]{#2}
\providecommand{\eprint}[2][]{\url{#2}}

\bibitem[{\citenamefont{Pudalov et~al.}()\citenamefont{Pudalov, Gershenson,
  Kojima, Butch, Dizhur, Brunthaler, Prinz, and Bauer}}]{exp}
\bibinfo{author}{\bibfnamefont{V.~M.} \bibnamefont{Pudalov}},
  \bibinfo{author}{\bibfnamefont{M.~E.} \bibnamefont{Gershenson}},
  \bibinfo{author}{\bibfnamefont{H.}~\bibnamefont{Kojima}},
  \bibinfo{author}{\bibfnamefont{N.}~\bibnamefont{Butch}},
  \bibinfo{author}{\bibfnamefont{E.~M.} \bibnamefont{Dizhur}},
  \bibinfo{author}{\bibfnamefont{G.}~\bibnamefont{Brunthaler}},
  \bibinfo{author}{\bibfnamefont{A.}~\bibnamefont{Prinz}}, \bibnamefont{and}
  \bibinfo{author}{\bibfnamefont{G.}~\bibnamefont{Bauer}},
  \bibinfo{howpublished}{\prl \textbf{88}, 196404 (2002); A. A. Shashkin, S. V.
  Kravchenko, V. T. Dolgopolov and T. M. Klapwijk, \prb \textbf{66}, 073303
  (2002); A. A. Shashkin, M. Rahimi, S. Anissimova, S. V. Kravchenko, V. T.
  Dolgopolov and T. M. Klapwijk, \prl \textbf{91}, 046403 (2003)}.

\bibitem[{\citenamefont{Smith and Stiles}()}]{oldexp}
\bibinfo{author}{\bibfnamefont{J.~L.} \bibnamefont{Smith}} \bibnamefont{and}
  \bibinfo{author}{\bibfnamefont{P.~J.} \bibnamefont{Stiles}},
  \bibinfo{howpublished}{\prl \textbf{29}, 102 (1972); P. T. Coleridge, M.
  Hayne, P. Zawadzki, A. S. Sachrajda, Surf. Sci. \textbf{361-361}, 560 (1996);
  W. Pan, D. C. Tsui and B. L. Draper, \prb \textbf{59}, 10208 (1999)}.

\bibitem[{\citenamefont{Rice}(1965)}]{rice}
\bibinfo{author}{\bibfnamefont{T.~M.} \bibnamefont{Rice}},
  \bibinfo{journal}{Ann. Phys. (N. Y.)} \textbf{\bibinfo{volume}{31}},
  \bibinfo{pages}{100} (\bibinfo{year}{1965}).

\bibitem[{\citenamefont{Gell-Mann}(1957)}]{gellmann}
\bibinfo{author}{\bibfnamefont{M.}~\bibnamefont{Gell-Mann}},
  \bibinfo{journal}{\prb} \textbf{\bibinfo{volume}{106}}, \bibinfo{pages}{369}
  (\bibinfo{year}{1957}).

\bibitem[{\citenamefont{Ting et~al.}(1975)\citenamefont{Ting, Lee, and
  Quinn}}]{ting}
\bibinfo{author}{\bibfnamefont{C.~S.} \bibnamefont{Ting}},
  \bibinfo{author}{\bibfnamefont{T.~K.} \bibnamefont{Lee}}, \bibnamefont{and}
  \bibinfo{author}{\bibfnamefont{J.~J.} \bibnamefont{Quinn}},
  \bibinfo{journal}{\prl} \textbf{\bibinfo{volume}{34}}, \bibinfo{pages}{870}
  (\bibinfo{year}{1975}).

\bibitem[{\citenamefont{Vinter}(1975)}]{vinter}
\bibinfo{author}{\bibfnamefont{B.}~\bibnamefont{Vinter}},
  \bibinfo{journal}{\prl} \textbf{\bibinfo{volume}{35}}, \bibinfo{pages}{1044}
  (\bibinfo{year}{1975}).

\bibitem[{\citenamefont{Abrikosov et~al.}()\citenamefont{Abrikosov, Gor'kov,
  and Dzyaloshinski}}]{book}
\bibinfo{author}{\bibfnamefont{A.~A.} \bibnamefont{Abrikosov}},
  \bibinfo{author}{\bibfnamefont{L.~P.} \bibnamefont{Gor'kov}},
  \bibnamefont{and} \bibinfo{author}{\bibfnamefont{I.~E.}
  \bibnamefont{Dzyaloshinski}}, \bibinfo{howpublished}{\textit{Methods of
  quantum field theory in statistical physics} (Dover Publications, New York,
  1963); G. D. Mahan, \textit{Many-Particle Physics} (Plenum Press, New York,
  1981); A. L. Fetter and J. D. Walecka, \textit{Quantum theory of
  many-particle systems} (McGraw-Hill, San Francisco, 1971)}.

\bibitem[{\citenamefont{Jalabert and {Das Sarma}}(1989)}]{jalabert}
\bibinfo{author}{\bibfnamefont{R.}~\bibnamefont{Jalabert}} \bibnamefont{and}
  \bibinfo{author}{\bibfnamefont{S.}~\bibnamefont{{Das Sarma}}},
  \bibinfo{journal}{\prb} \textbf{\bibinfo{volume}{40}}, \bibinfo{pages}{9723}
  (\bibinfo{year}{1989}).

\bibitem[{\citenamefont{Marmorkos and {Das Sarma}}(1991)}]{marmorkos}
\bibinfo{author}{\bibfnamefont{I.~K.} \bibnamefont{Marmorkos}}
  \bibnamefont{and} \bibinfo{author}{\bibfnamefont{S.}~\bibnamefont{{Das
  Sarma}}}, \bibinfo{journal}{\prb} \textbf{\bibinfo{volume}{44}},
  \bibinfo{pages}{3451} (\bibinfo{year}{1991}).

\bibitem[{\citenamefont{Chaplik}(1971)}]{chaplik}
\bibinfo{author}{\bibfnamefont{A.~V.} \bibnamefont{Chaplik}},
  \bibinfo{journal}{Sov. Phys. - JETP} \textbf{\bibinfo{volume}{33}},
  \bibinfo{pages}{997} (\bibinfo{year}{1971}).

\bibitem[{\citenamefont{Hu}()}]{hu}
\bibinfo{author}{\bibfnamefont{B.~Y.~K.} \bibnamefont{Hu}},
  \bibinfo{howpublished}{\prb \textbf{47}, 1687 (1993); B. Y. K. Hu and S. {Das
  Sarma}, \prb, \textbf{48}, 5469 (1993)}.

\bibitem[{\citenamefont{{Das Sarma} et~al.}({\natexlab{a}})\citenamefont{{Das
  Sarma}, Kalia, Nakayama, and Quinn}}]{dassarma}
\bibinfo{author}{\bibfnamefont{S.}~\bibnamefont{{Das Sarma}}},
  \bibinfo{author}{\bibfnamefont{P.~K.} \bibnamefont{Kalia}},
  \bibinfo{author}{\bibfnamefont{M.}~\bibnamefont{Nakayama}}, \bibnamefont{and}
  \bibinfo{author}{\bibfnamefont{J.~J.} \bibnamefont{Quinn}},
  \bibinfo{howpublished}{\prb \textbf{19}, 6397 (1979); S. Das Sarma and B.
  Vinter, \prb \textbf{28}, 3639 (1983)}.

\bibitem[{\citenamefont{Giuliani and Quinn}()}]{gq}
\bibinfo{author}{\bibfnamefont{G.~F.} \bibnamefont{Giuliani}} \bibnamefont{and}
  \bibinfo{author}{\bibfnamefont{J.~J.} \bibnamefont{Quinn}},
  \bibinfo{howpublished}{\prb \textbf{26}, 4421 (1982); L. Zheng and S. Das
  Sarma, \prb \textbf{53}, 9964 (1996)}.

\bibitem[{\citenamefont{Chubukov and Maslov}()}]{chubukov}
\bibinfo{author}{\bibfnamefont{A.~V.} \bibnamefont{Chubukov}} \bibnamefont{and}
  \bibinfo{author}{\bibfnamefont{D.~L.} \bibnamefont{Maslov}},
  \eprint{cond-mat/0304381; cond-mat/0305022}.

\bibitem[{\citenamefont{{Das Sarma} et~al.}({\natexlab{b}})\citenamefont{{Das
  Sarma}, Galitski, and Zhang}}]{short}
\bibinfo{author}{\bibfnamefont{S.}~\bibnamefont{{Das Sarma}}},
  \bibinfo{author}{\bibfnamefont{V.~M.} \bibnamefont{Galitski}},
  \bibnamefont{and} \bibinfo{author}{\bibfnamefont{Y.}~\bibnamefont{Zhang}},
  \eprint{cond-mat/0311559}.

\bibitem[{\citenamefont{Zhang and {Das Sarma}}()}]{diverge}
\bibinfo{author}{\bibfnamefont{Y.}~\bibnamefont{Zhang}} \bibnamefont{and}
  \bibinfo{author}{\bibfnamefont{S.}~\bibnamefont{{Das Sarma}}},
  \eprint{cond-mat/0312565}.

\bibitem[{\citenamefont{Lundqvist}()}]{lundqvist}
\bibinfo{author}{\bibfnamefont{B.~I.} \bibnamefont{Lundqvist}},
  \bibinfo{howpublished}{Phys. Kondens. Mater. \textbf{6}, 206 (1967); A. W.
  Overhauser, \prb \textbf{18} 2884 (1978)}.

\bibitem[{\citenamefont{Ando et~al.}(1982)\citenamefont{Ando, Fowler, and
  Stern}}]{ando}
\bibinfo{author}{\bibfnamefont{T.}~\bibnamefont{Ando}},
  \bibinfo{author}{\bibfnamefont{A.~B.} \bibnamefont{Fowler}},
  \bibnamefont{and} \bibinfo{author}{\bibfnamefont{F.}~\bibnamefont{Stern}},
  \bibinfo{journal}{Rev. Mod. Phys.} \textbf{\bibinfo{volume}{54}},
  \bibinfo{pages}{437} (\bibinfo{year}{1982}).

\bibitem[{\citenamefont{Galitski and {Das Sarma}}()}]{vd}
\bibinfo{author}{\bibfnamefont{V.~M.} \bibnamefont{Galitski}} \bibnamefont{and}
  \bibinfo{author}{\bibfnamefont{S.}~\bibnamefont{{Das Sarma}}},
  \eprint{cond-mat/03033632}.

\end{thebibliography}

\end{document}